\documentclass{ws-procs9x6}

\def\etal {{\it et al.}}

\def\ga{\gamma}

\def\om{\omega}

\def\cl{{\cal L}}

\def\fr#1#2{{{#1} \over {#2}}}
\def\half{{\textstyle{1\over 2}}}
\def\frac#1#2{{\textstyle{{#1}\over {#2}}}}

\def\lsim{\mathrel{\rlap{\lower4pt\hbox{\hskip1pt$\sim$}}
    \raise1pt\hbox{$<$}}}
\def\gsim{\mathrel{\rlap{\lower4pt\hbox{\hskip1pt$\sim$}}
    \raise1pt\hbox{$>$}}}

\def\ol#1{\overline{#1}}

\def\etal{{\it et al.}}

\def\lrpartial{\raise 1pt\hbox{$\stackrel\leftrightarrow\partial$}}

\def\lrprtmu{\stackrel{\leftrightarrow}{\partial_\mu}}
\def\lrprtnu{\stackrel{\leftrightarrow}{\partial^\nu}}

\newcommand{\beq}{\begin{equation}}
\newcommand{\eeq}{\end{equation}}
\newcommand{\bea}{\begin{eqnarray}}
\newcommand{\eea}{\end{eqnarray}}

\begin{document}

\title{LORENTZ VIOLATION IN TOP-QUARK PHYSICS}

\author{M.S.\ BERGER}

\address{Department of Physics, Indiana University\\
Bloomington, IN, 47405, USA\\
E-mail: berger@indiana.edu}

\begin{abstract}
Lorentz and CPT violation can affect the rates for $t$-$\ol t$ production and decay.
The Lorentz-violating coefficients in the Standard-Model Extension responsible
for modifying the top-quark events have recently been bounded by the D0 Collaboration. 
To extend the analysis to the LHC the calculations need to be extended to include
the gluon fusion production mechanism. Some of the first results of this program were 
presented at the Meeting. 
\end{abstract}

\bodymatter

\section{Introduction}

The possible existence of Lorentz and CPT violation has been increasingly studied in recent years. 
Many efforts toward developing a theory of quantum gravity contain Lorentz violation as 
a generic physical consequence although the size of the effects are usually quite small in realistic
experiments with attainable energies much smaller than the Planck scale.
A systematic approach for parametrizing the possible forms of Lorentz violation in the context 
of an effective field theory called the Standard-Model Extension (SME)\cite{dcak1,dcak2,ak} 
has been developed. The Lorentz- and CPT-violating physical effects are encoded in the form of coefficients. 
A Lorentz-violating term is an observer scalar density formed by contracting the Lorentz-violating 
operator with the associated coefficient. Diverse kinds of experiments can place bounds on the 
same coefficients, so that the experimental sensitivities can be directly compared\cite{cptprocs}.

The Lorentz-violating coefficients provide the possibility of preferred frames and directions
in space and could in principle have an effect on collider physics experiments. If the coefficient
is constant in space then one expects there to be a sidereal time variation as the Earth 
rotates through this constant background. Collider physics scattering events naturally have access to 
many Lorentz frames since produced particles and their decay products 
are distributed with many different boosts and orientations relative to the laboratory frame and beam
direction. 

\section{Top quark Lorentz-violating coefficients in the SME}

Top quark production and decay involves many fields so the overall rates and the distributions
could be affected by 
many different Lorentz violating coefficients in the SME. In this discussion we will focus
on coefficients which involve the top quark field $t$. Then the  relevant coefficients arise
from the matrices 
$(a_Q)_{\mu ij}$, 
$(a_U)_{\mu ij}$, 
$(c_Q)_{\mu\nu ij}$, 
$(c_U)_{\mu\nu ij}$, 
$(H_U)_{\mu\nu ij}$
contained in the SME,
where $i$ and $j$ are flavor indices, and the $Q$ and $U$ indices 
correspond to the left- and right-handed fields of the Standard Model. 
Of these five coefficients, the first two control CPT-odd operators,
while the last three control CPT-even ones. At present bounds have been 
placed on the $c$-type coefficients\cite{Abazov:2012iu}, so we will 
concentrate on those in the following.
The coefficients relevant to the top quark production and decay have $i=j=3$ 
(in general there will be
flavor mixing in the Lorentz-violation sector that does not have to align
with the CKM matrix of the Standard Model), so
for convenience one can define 
$(c_L)_{\mu\nu}
=
(c_Q)_{\mu\nu 33}$
and
$(c_R)_{\mu\nu}
=
(c_U)_{\mu\nu 33}$
to simplify the notation. 

The Standard-Model terms involving 
the $t$ and $b$ quark fields
and their interactions with the $W^\pm_\mu$ boson
are 
\bea
\cl_{t,b}^{\rm SM} &=& 
\half i \ol{t} \ga^\mu \lrprtmu t 
- m_t \ol t t+ \half i \ol{b} \ga^\mu \lrprtmu b 
- m_b \ol b b
\nonumber \\ &&
+ (\fr g {\sqrt{2}} W_\mu^- \ol b_L \ga^\mu t_L + \rm{h.c.}).
\label{smquarks}
\eea

The additional terms in the SME corrections involving the 
$c$-type coefficient
can be written in equivalent forms,
\bea
\cl^{\rm CPT-even}_{t,b} &=& 
\half i (c_L)_{\mu\nu} \ol t_L \ga^\mu \lrprtnu t_L 
+ \half i (c_R)_{\mu\nu} \ol t_R \ga^\mu \lrprtnu t_R 
\nonumber \\ &&
+ \half i (c_L)_{\mu\nu} \ol b_L \ga^\mu \lrprtnu b_L
+ (\fr g {\sqrt{2}} (c_L)_{\mu\nu} W^{-\nu} 
\ol b_L \ga^\mu t_L + \rm{h.c.})
\nonumber \\ 
&=&
\half i c_{\mu\nu} \ol t \ga^\mu \lrprtnu t 
+ \half i d_{\mu\nu} \ol t \ga^5 \ga^\mu \lrprtnu t 
\nonumber \\ &&
+ \half i (c_L)_{\mu\nu} \ol b_L \ga^\mu \lrprtnu b_L
+ (\fr g {\sqrt{2}} (c_L)_{\mu\nu} W^{-\nu} 
\ol b_L \ga^\mu t_L + \rm{h.c.}),
\qquad
\label{loryukawa}
\eea
where $c_{\mu\nu}= (c_L)_{\mu\nu}+(c_R)_{\mu\nu}$ and 
$d_{\mu\nu}= (c_L)_{\mu\nu}-(c_R)_{\mu\nu}$. 
These terms give rise to additional Feynman rules which can be treated 
as insertions in diagrams. By working in leading order in the 
Lorentz-violating coefficients, it is a straightforward procedure to enumerate
diagrams and calculate contributions to the amplitudes which depend on 
them. At the Tevatron top quark production is dominated by the $q\bar{q}\to t\bar{t}$
process for which the Lorentz-violating corrections are known.

\section{Bounds}

The Lorentz violating coefficients are defined as constant in a Sun-centered 
frame\cite{datatables,akmm,space} ($X$, $Y$, $Z$, $T$).
The coefficients in the 
laboratory frame ($x$, $y$, $z$, $t$) then change with sidereal time because
the Earth is rotating. Nonzero Lorentz-violating coefficients thus
produce sidereal 
time-dependent signals with frequency an integer multiple
of $\om_\oplus\simeq 2\pi$/(23 h 56 min.).\cite{aksidereal}
The D0 Collaboration has used data collected at the Fermilab Tevatron Collider from 5.3~fb$^{-1}$ of 
integrated luminosity to place the
first bounds on the top-quark Lorentz-violating coefficients\cite{Abazov:2012iu,whittington}: 
\begin{eqnarray}
&&(c_Q)_{XX33}=-0.12\pm 0.11\pm 0.02,
\nonumber \\
&&(c_Q)_{YY33}=0.12\pm 0.11\pm 0.02,
\nonumber \\
&&(c_Q)_{XY33}=-0.04\pm 0.11\pm 0.01,
\nonumber \\
&&(c_Q)_{XZ33}=0.15\pm 0.08\pm 0.02,
\nonumber \\
&&(c_Q)_{YZ33}=-0.03\pm 0.08\pm 0.01,
\nonumber \\
&&(c_U)_{XX33}=0.1\pm 0.09\pm 0.02,
\nonumber \\
&&(c_U)_{YY33}=-0.1\pm 0.09\pm 0.02,
\nonumber \\
&&(c_U)_{XY33}=0.04\pm 0.09\pm 0.01,
\nonumber \\
&&(c_U)_{XZ33}=-0.14\pm 0.07\pm 0.02,
\nonumber \\
&&(c_U)_{YZ33}=0.01\pm 0.07\pm <0.01.
\end{eqnarray}
The experimental events used in the analysis contain lepton $+$ jets final states, and a candidate signal 
would be expected to exhibit sidereal or 
twice-sidereal variations. These bounds are at roughly the 10\% level, and all coefficients are consistent with zero.
It is expected that the much larger top sample at the LHC could improve these bounds by
at least an order of magnitude.

\section{Gluon Fusion}

Top-quark production is dominated by the gluon-fusion mechanism at the LHC.  There are three
contributing diagrams ($s$-channel gluon and $t$- and $u$- channel top 
quark) in the Standard Model, and the cross section is 
well-known\cite{Gluck:1977,Combridge:1978kx,Ellis:1991qj}. The calculation of the Lorentz-violating corrections is
somewhat more involved than the one for $q\bar{q}\to t\bar{t}$
because of the number of diagrams and the presence of external gluons which must be
taken to be transverse. The first results of this calculation were presented at the Meeting. 

\section{Summary}

The first bounds on the top-quark SME coefficients have been obtained by
the D0 Collaboration using data collected at the Tevatron. The measurements 
are consistent with zero. Top quark production at the LHC is dominated by 
gluon fusion, and it is expected that the bounds can be improved to the percent level. 

\section*{Acknowledgments}

This work is supported in part
by DOE grant DE-FG02-91ER40661 and by the Indiana University
Center for Spacetime Symmetries. The author thanks V.A.\ Kosteleck\'y for his 
collaboration.


\begin{thebibliography}{xx}

\bibitem{dcak1}
D.\ Colladay and V.A.\ Kosteleck\'y,
Phys.\ Rev.\ D {\bf 55}, 6760 (1997).

\bibitem{dcak2}
D.\ Colladay and V.A.\ Kosteleck\'y,
Phys.\ Rev.\ D {\bf 58}, 116002 (1998).

\bibitem{ak}
V.A.\ Kosteleck\'y,
Phys.\ Rev.\ D {\bf 69}, 105009 (2004).

\bibitem{cptprocs}
V.A.\ Kosteleck\'y, ed.,
{\it CPT and Lorentz Symmetry I, II, III, IV, V},
World Scientific, Singapore,
1999, 2002, 2005, 2008, 2011.

\bibitem{Abazov:2012iu} 
  V.M.\ Abazov {\it et al.},
  Phys.\ Rev.\ Lett.\  {\bf 108}, 261603 (2012).

\bibitem{akmm}
V.A.\ Kosteleck\'y and M.\ Mewes,
Phys.\ Rev.\ D {\bf 66}, 056005 (2002).

\bibitem{space}
R.\ Bluhm \etal,
Phys.\ Rev.\ D {\bf 68}, 125008 (2003).

\bibitem{datatables}
{\it Data Tables for Lorentz and CPT Violation,}
V.A.\ Kosteleck\'y and N.\ Russell,
2013 edition,
arXiv:0801.0287v6.

\bibitem{aksidereal}
V.A.\ Kosteleck\'y,
Phys.\ Rev.\ Lett.\ {\bf 80}, 1818 (1998).

\bibitem{whittington} D.\ Whittington, these proceedings.

\bibitem{Gluck:1977} 
  M.\ Gluck, J.F.\ Owens and E.\ Reya,
  Phys.\ Rev.\ D {\bf 17}, 2324 (1978).

\bibitem{Combridge:1978kx} 
  B.L.\ Combridge,
  Nucl.\ Phys.\ B {\bf 151}, 429 (1979).

\bibitem{Ellis:1991qj} 
  R.K.\ Ellis, W.J.\ Stirling and B.R.\ Webber,
  Camb.\ Monogr.\ Part.\ Phys.\ Nucl.\ Phys.\ Cosmol.\  {\bf 8}, 1 (1996).

\end{thebibliography}
\end{document}